\documentclass[final,english]{bullsrsl}[2022/06/15]

\usepackage[latin1]{inputenc}
\usepackage[T1]{fontenc}
\usepackage{multirow}
\usepackage{natbib} 
\usepackage{graphicx}
\usepackage{subcaption}
\usepackage{graphicx}	
\usepackage{amsmath}	
\usepackage{amssymb}	
\usepackage{float}
\usepackage{natbib}
\usepackage{morefloats}
\usepackage{dcolumn}
\usepackage{amsmath}
\usepackage{latexsym}
\usepackage{longtable}
\usepackage{lipsum}
\usepackage{gensymb} 
\usepackage{adjustbox}
\usepackage{lscape}
\usepackage{empheq}
\usepackage{mathrsfs}
\usepackage{textcomp}
\usepackage{gensymb}
\usepackage{hyperref}
\usepackage{graphicx}
\usepackage{multirow}
\usepackage{longtable}
\usepackage{booktabs}

\begin{document}
\title{Prompt emission properties of GRB~200613A}

\author[affil={1}, corresponding]{Ankur}{Ghosh}
\author[affil={1}]{Kuntal}{Misra}
\author[affil={1,2}]{Dimple}{}

\affiliation[1]{Aryabhatta Research Institute of observational sciences, Manora Peak, Nainital 263 001, India}
\affiliation[2]{Department of Physics, Deen Dayal Upadhyaya Gorakhpur University, Gorakhpur-273009, India}

\correspondance{ghosh.ankur1994@gmail.com}
\date{13th October 2020}
\maketitle


%

\begin{abstract}

We study the prompt emission properties of the long duration GRB~200613A using \textit{Fermi}-Gamma-Ray Burst Monitor (GBM) and Large Area Telescope (LAT) data. The prompt emission light curve of GRB~200613A reveals a strong peak emission up to $\sim$ 50 s after the burst accompanied by an extended emission up to $\sim$ 470 s similar to that seen in ultra-long GRB light curves. The time-integrated spectroscopy shows that the Band function best fits the main emission episode, and the extended emission follows the power-law behaviour because of poor count rates. Due to its high isotropic energy and low peak energy, GRB~200613A lies at the extreme end in both the $E_{\rm p}$--$E_{\rm iso}$ and $E_{\rm p}$--$T_{90}$ plots. In addition to the GBM detection, the \textit{Fermi}-LAT detected the highest energetic photons of 7.56 GeV after 6.2 ks since burst, which lies beyond the maximum synchrotron energy range.
\end{abstract}


\section{Introduction}
\label{ch1:intro}

Gamma-Ray Bursts (GRBs) are extremely energetic transient events in the Universe, releasing colossal amounts of energy which can last from a few milliseconds to several thousand seconds. Traditionally, GRBs are classified into two sub-categories (long and short) based on their $T_{90}$ duration (the time interval during which 90\% of the total fluence has been estimated; \citealt{1993ApJ...413L.101K}). Although this duration-based classification is sensitive to many factors, it is still valid to a certain extent. In the last decade, a subset of GRBs was detected with duration $>$ 500 s, known as ultra-long GRBs \citep{2013ApJ...779...66S, 2014ApJ...781...13L}. The unique and relatively rare sub-class of GRBs typically consists of a shorter main episode and an extended quiescent emission, attributed to late-time central engine activity \citep{2014ApJ...787...66Z}. Their extended duration far exceeds typical long GRBs, posing intriguing questions about the physical mechanisms responsible for such long-lasting bursts \citep{2015ApJ...800...16B}.

Although multiple empirical models such as Band function \citep{1997ApJ...486..928B}, cut-off power-law (CPL), smoothly-broken power-law (SBPL), and black body (BB) can explain the prompt emission mechanism in GRBs but it is still poorly understood. Recent studies by  \citet{2020NatAs...4..174B} and \citet{ 2020ApJ...893..128A} have shown that physical models like synchrotron and photospheric emission could be the primary components of prompt emission. With the detection of ultra-long GRBs, it is vital to study the prompt emission mechanism and the characteristics of the long-lived central engine. 

GRB~200613A (z = 1.22) lies at the boundary between long and ultra-long sub-classes with the burst duration $\sim$ 470 s. It has a strong emission component up to $\sim$50 s and an extended emission (from 200 s to 470 s having a signal-to-noise ratio of 3.5$\sigma$), almost resembling the temporal structure of ultra-long GRBs. This study presents a comprehensive prompt emission study of GRB~200613A using the {\it Fermi} data and compares it with other GRBs. The observations and data reduction are briefly described in section \ref{ch2:reduction}. The results obtained are presented in section \ref{ch3:analysis}. A concise summary of this work is presented in section \ref{ch4:sum}. We assume flat cosmology throughout the paper with $H_0 = 71$ $\rm km~s^{-1} Mpc^{-1}$, $\Omega_m$ = 0.27, and $\Omega_{\lambda}$ = 0.73 \citep{Komatsu2011}.

\section{ \textit{Fermi} data reduction and analysis}
\label{ch2:reduction}

\subsection{GBM light curve and spectra}
\label{lcspec}

\begin{table}

\caption{Properties of GRB~200613A}  

\label{200613A_prop}
\begin{center}

\begin{tabular}{ |c||c||c| } 
\hline
Parameters & Values & References \\
\hline\hline
$T_{90}$ duration (s) &  470 & \citet{2020GCN.27930....1B}\\
Redshift & 1.22 & \citet{2021GCN.29320....1D}\\
Fluence (erg $cm^{-2}$)  &   (4.10 $\pm$ 0.05) $\times 10^{-5}$ & \citet{2020GCN.27930....1B} \\
$E_{\rm iso}$ (erg) & $1.37_{-0.04}^{+0.05} \times 10^{53}$ & This work\\
$E_{\rm p}$ (keV) &  115 $\pm$ 5  	& This work\\

\hline
\end{tabular}

\end{center}

\end{table}

\begin{figure}
\centering
\includegraphics[width=0.6\textwidth]{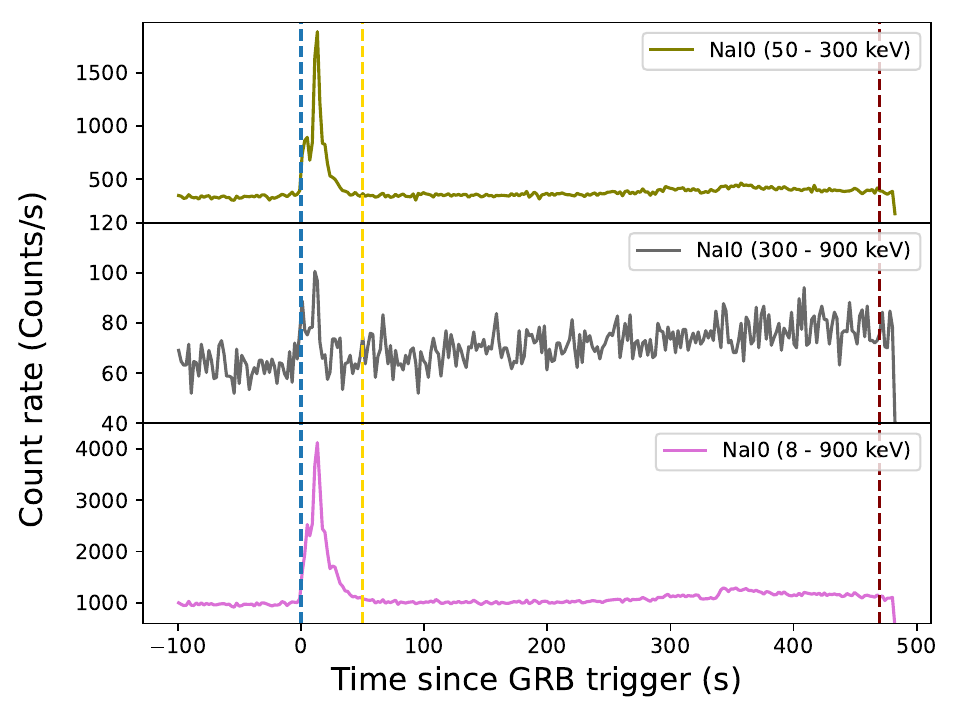}
\begin{minipage}{12cm}
\caption{Multichannel light curve of GRB~200613A of NaI0 detector in three different energy bins (50-300 keV, 300-900 keV, and 8-900 keV). The dashed vertical gold line indicates the strongest peak emission up to $50$ s. $T_{90}$ duration of the GRB is denoted with the maroon dashed line.}
\label{fig:lc}
\end{minipage}
\end{figure}

GRB~200613A was triggered and located by \textit{Fermi}-GBM at 05:30:08 UT on 13 Jun 2020 \citep{2020GCN.27926....1F}. \textit{Fermi}-LAT detected the GRB at the same time as the boresight angle at the time of GBM trigger was $25^{\circ}$ \citep{2020GCN.27931....1O}. The GBM data was taken from the public data archive of \textit{Fermi} satellite. We reduced the GBM data following the standard process of \texttt{GBM Data Tools} and \texttt{GTBurst}. We picked two NaI0 and NaI1 detectors for our further analysis as the count rates are maximum and observing angles are minimum for those detectors. The angle constraint is imposed because the systematic uncertainty at larger angles becomes significant. The BGO0 detector was chosen over BGO1 because of the same reason. Fig. \ref{fig:lc} shows the GBM light curves in (50-300) keV, (300-900) keV, and (8-900) keV energy ranges. The prompt emission light curve in (50-300) keV exhibits a bright emission from $T_0$ to $T_{0} + 50$ s, followed by a comparatively weaker emission up to 470 s since the explosion. The weaker extended emission is not evident in the (300-900) keV light curve.

To investigate the radiation mechanism associated with the prompt emission of GRB~200613A, we performed time-integrated spectral analysis using threeML \citep{2015arXiv150708343V}. The spectral analysis was carried out for the NaI (8-900 keV) and BGO (200-30000 keV) detectors. For fitting the bright phase of the GRB (up to 50 s since the burst), multiple phenomenological models such as Band function, CPL, SBPL, and PL+BB were considered. We adopted the Bayesian parameter estimation technique using the `dynesty-nested' sampler in threeML to estimate the best-fit value.
Depending on the Akaike Information Criteria (AIC), Bayesian Information Criteria (BIC; Kass \& Rafferty 1995), and Log(likelihood) for individual models, the Band function is the best-suited model for this period. The best-fitted parameters of each model are given in Table \ref{tab:gbm_spec}. Due to the poor count rate in the extended emission region, only the power law model can provide a better fit to the spectra.

\begin{table}

\caption{Results of spectral analysis of GRB~200613A based on {\it Fermi} - GBM data.}  

\label{tab:gbm_spec}
\begin{tabular}{| p{1.75cm}|p{2.5cm}||p{1.75cm}|p{2.5cm}||p{2.0cm}| p{2.75cm}|}
 \hline
 \multicolumn{6}{|c|}{The list of model parameters for peak interval (9 - 17 s)} \\
 \hline
  Band      & Best fit value &  CPL & Best fit value & SBPL  & Best fit value \\
  parameters&                &  parameters &         &parameters &\\
 \hline
 $\alpha$   &  $-0.95_{-0.02}^{+0.02}$  &  index &   $-0.96_{-0.02}^{+0.02}$       &  $\alpha$    & $-1.07_{-0.02}^{+0.02}$\\
 $E_p (keV)^{1}$&  $+151.30_{-3.20}^{+3.00}$       &  $E_c (keV)^{2}$ &  $+148.00_{-5.00}^{+6.00}$       &  $E_b (keV)^{3}$ & $+223.00_{-19.00}^{+21.00}$\\
 $\beta$    &  $-3.79_{-0.32}^{+0.42}$     &        &             &  $\beta$     & $-4.01_{-0.28}^{+0.23}$\\
 BIC        &  $+3749.41$                     & BIC    &  $+3778.81$    & BIC          & $+3755.25$           \\
 $Log(Z)^{4}$     &  $-811.86$                    & log(Z) &  $-813.05$   &log(Z)        & $-812.82$\\
 
 \hline

\end{tabular}
 \footnotesize{$^{1}$ Peak energy, $^2$ Cutoff energy, $^3$ Break energy, $^4$ Evidence}

\end{table}

\subsection{LAT data}
\label{spectra}

\begin{figure}
\centering
\includegraphics[width=0.5\textwidth]{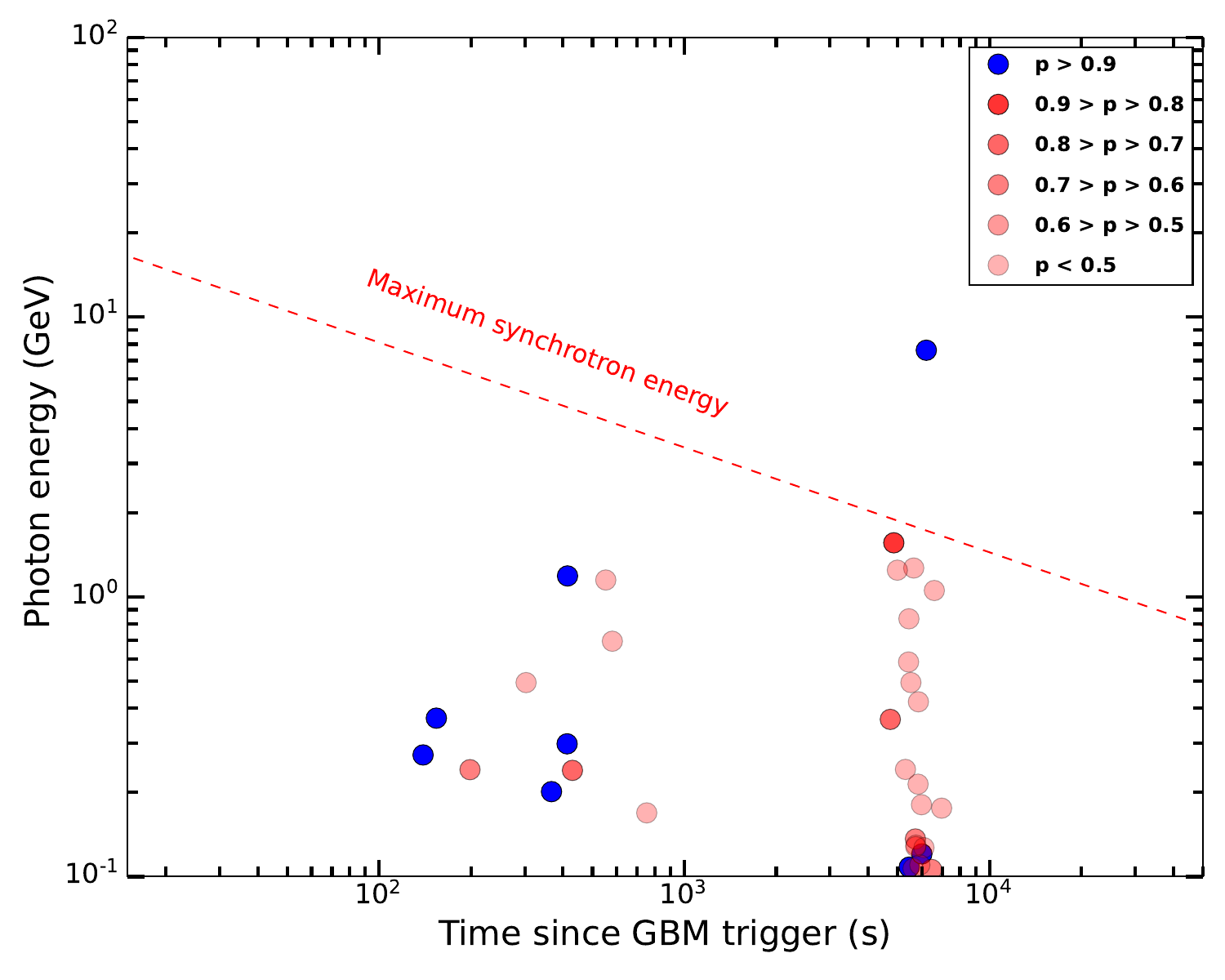}
\begin{minipage}{12cm}
\caption{Photons detected by {\it Fermi} - LAT with their respective energies and probabilities of being associated with GRB~200623A. The maximum energy photon (with $>$ 90\% probability) detected by LAT is 7.56 GeV. The diagonal dashed line represents the theoretical limit of maximum synchrotron energy of electron}
\label{fig:lat}
\end{minipage}
\end{figure}

The LAT data was extracted up to $5 \times 10^4$ s since the burst from the LAT catalogue. We carried out unbinned likelihood analysis using \texttt{GTBurst} software. The energy bin selected for the LAT analysis is 100 MeV - 300 GeV, and the maximum zenith angle is chosen $100^{\circ}$ to discard photons coming from Earth's limb. As GRB~200613A is a long duration event ($> 10^{3}$ s), P8R3\textunderscore SOURCE\textunderscore V3 was taken as a response. We use \texttt{gtsrcprob} task for checking the probability of photons associated with the GRB~200613A. The maximum energetic photon detected by LAT for this event is 7.56 GeV, observed at 6.2 ks since the burst with the probability of association with the source is $>$ 90\%. To investigate the origin of the LAT detected photons, we plotted the maximum synchrotron energy possible by photons with the dashed red line in Fig.\ref{fig:lat} \citep{2010ApJ...718L..63P, 2019ApJ...879L..26F}. All the photons of GRB~200613A detected by LAT are lying below the line following the synchrotron emission except for the maximum energetic photon. This indicates that the 7.56 GeV photon comes from a separate origin.   

\section{Location of GRB~200613A in the $E_{\rm p}$--$E_{\rm iso}$ and $E_{\rm p}$--$T_{90}$ plots}
\label{ch3:analysis}

\subsection{$E_{\rm p}$--$E_{\rm iso}$ Correlation}
\label{amati}

\begin{figure}
     \centering
     \begin{subfigure}[b]{0.48\textwidth}
         \centering
         \includegraphics[width=\linewidth]{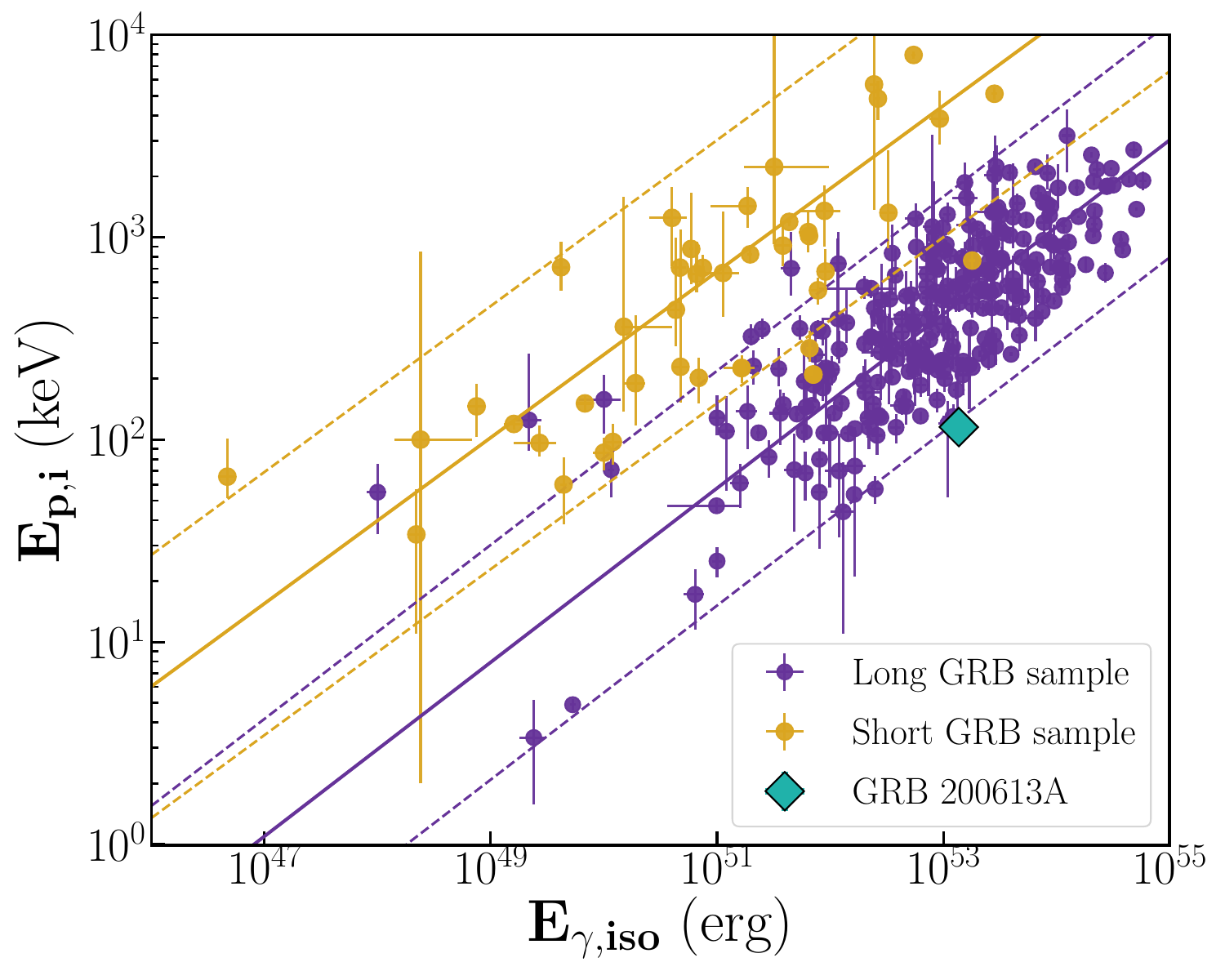}
         \caption{$E_{\rm p}$--$E_{\rm iso}$ correlation}
         \label{fig:amati}
     \end{subfigure}
     \hfill
     \begin{subfigure}[b]{0.48\textwidth}
         \centering
         \includegraphics[width=\linewidth]{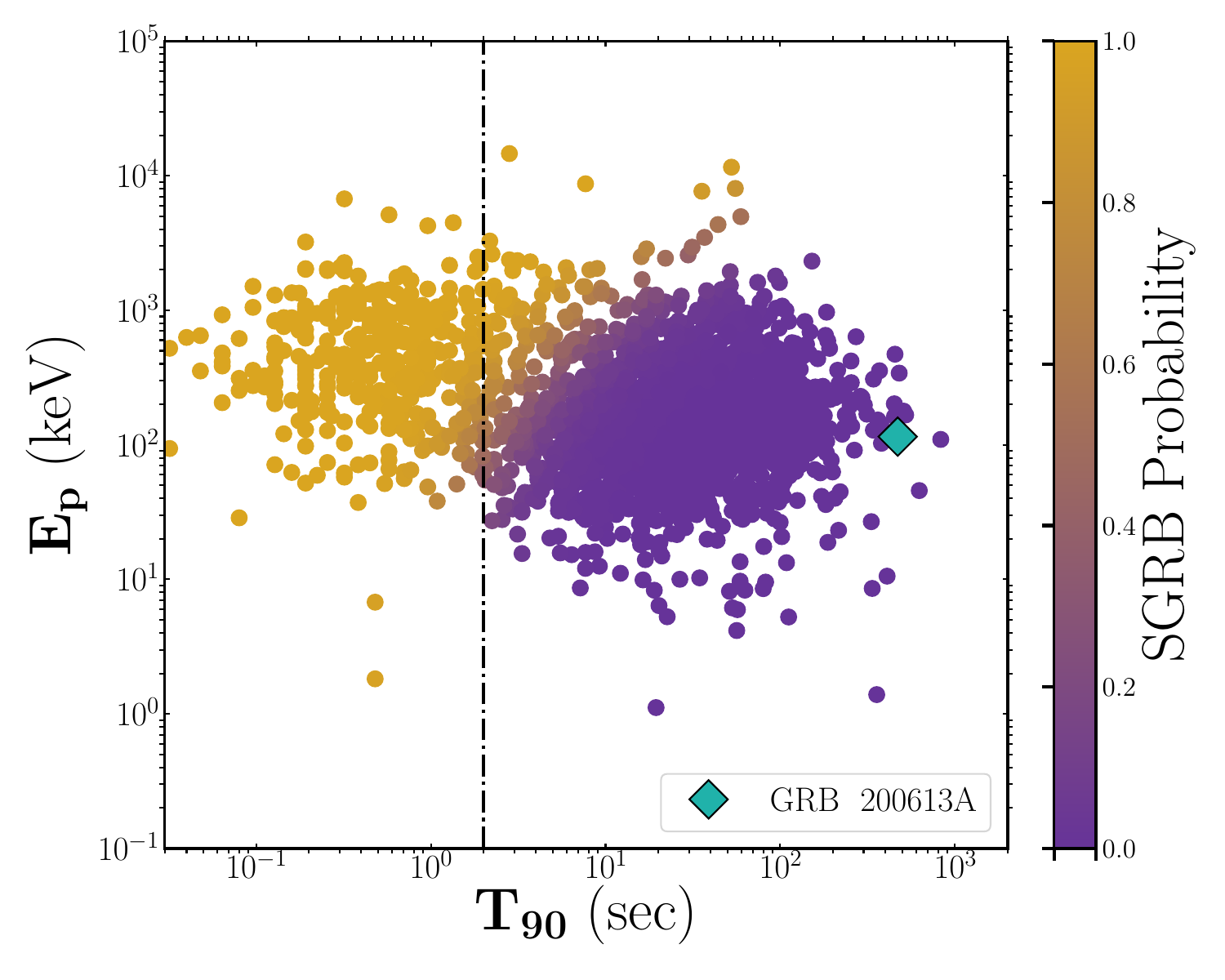}
         \caption{$E_p - T_{90}$ correlation}
         \label{fig:hardness}
     \end{subfigure}
     \hfill
      \begin{minipage}{12cm}   
        \caption{The $E_{\rm p}$--$E_{\rm iso}$ and $E_{\rm p}$--$T_{90}$ correlation for the sample of long (purple circles) and short (goldenrod circles) GRBs. In the $E_{\rm p}$--$E_{\rm iso}$ plot, the dashed line in the same colours corresponds to the 3-$\sigma$ scatter of the correlation. The light-green diamond symbol represents the position GRB~200613A in these plots.} 
     \end{minipage}   
        \label{fig:corr}
\end{figure}

A unique property of GRB prompt emission is that the spectral peak energy of the $\nu F_{\nu}$ spectrum ($E_{p,i}$) and the isotropic equivalent energy of the prompt emission are correlated \citep{2006MNRAS.372..233A}. This correlation is primarily used in prompt emission studies to classify the GRBs and estimate their redshifts if the spectroscopic redshifts are unknown. The correlation observed among different types of investigated gamma-ray transients is supported by empirical evidence indicating a power law relationship. Moreover, these transients exhibit distinct and well-separated regions. Fig. \ref{fig:amati} depicts the correlation between 316 short and long GRBs published in \citet{min21}. 

Using the time-integrated fluence value in 10-1000 keV and redshift information from the host galaxy observation, the calculated $E_{\gamma,iso}$ value is (1.37 $\pm$ 0.16) $\times 10^{53}$ erg. The spectral peak energy for GRB~200613A is $E_{p,i}$ is $114.90_{-3.00}^{+5.00}$ keV (Table \ref{200613A_prop}). From Fig. \ref{fig:amati}, it is evident that GRB~200613A lies beyond the 3-$\sigma$ boundary of the long GRBs.  

\subsection{$E_{\rm p}$--$T_{90}$ Correlation}
\label{sphardness}

Another method of classifying the GRBs is spectral hardness and peak energy ($E_{p}$) correlation. To calculate the probability of a GRB being short or long, we used the Bayesian Gaussian Mixture Model (BGMM). Fig. \ref{fig:hardness} shows a bimodal distribution between the peak energy and the $T_{90}$ duration of GRB prompt emission.

The hardness ratio (HR) can be interpreted as the ratio of counts between two energy bands. We considered 8-50 keV and 50-300 keV bands for our calculation. The calculated value of HR for GRB~200613A is $0.58 \pm 0.05$. The $E_{\rm p}$ is calculated via model fitting of GBM data, which comes to around $115 \pm 5$ keV (Table \ref{200613A_prop}). We estimated the $T_{90}$ duration of GRB~200613A considering the total emission episodes using GBM data that is $470 \pm 5$ s. From Fig. \ref{fig:hardness}, we see that GRB~200613A lies at the extreme end of the duration distribution.

\section{Summary}
\label{ch4:sum}

We carried out a comprehensive, prompt emission analysis of GRB~200613A where different parameters such as $T_{90}$, HR, $E_{\rm p}$, and $E_{iso}$ were estimated. The parameters were compared with a larger sample of GRBs. The burst duration of GRB~200613A suggests that it lies at the boundary of the long and ultra-long duration GRBs. The light curve in 50-300 keV consists of a bright peak emission up to $\sim$ 50 s followed by a weaker emission up to 470 s. However, the evidence of extended emission is not seen in the light curve of the 300-900 keV energy band. The temporal structure of GRB~200613A resembles the ultra-long GRBs. We have also performed time-integrated spectral analysis for the four GBM detectors (NaI0, NaI1, NaI6, BGO0) with the highest count rates, suggesting that the Band function can explain the peak emission (9 - 17 s). In contrast, the late-time weak extended emission best fits with a power law because of poor count rates. In both $E_{\rm p}$--$E_{\rm iso}$ and $E_{\rm p}$--$T_{90}$ correlations, this burst is located at the extreme end of the long GRB sample. The isotropic energy ($E_{\rm iso}$) of GRB~200613A estimated from the prompt emission study is $1.37_{-0.04}^{+0.05} \times 10^{53}$ erg, which enlists this GRB in the hyper-energetic GRBs detected by \textit{Fermi}. {\it Fermi}-LAT detected a 7.56 GeV photon 6.2 ks after the burst trigger, which lies above the maximum synchrotron energy of photons and might come from other origins such as proton synchrotron and synchrotron self-Compton emission \citep{2020MNRAS.496..974Z, 2023ApJ...947L..14Z}.

This study presents the prompt emission properties of GRB~200613A. We have yet to explore the afterglow properties of this burst. However, complete prompt emission and afterglow analysis of GRB~200613A will be useful to provide conclusive information about the progenitor channel, jet properties, emission mechanism, and environment, which will be carried out in future. 

\begin{acknowledgments}
We thank the referee for providing valuable comments which have improved the presentation and content of the manuscript.
\end{acknowledgments}

\begin{furtherinformation}

\begin{orcids}
\orcid{0000-0003-2265-0381}{Ankur}{Ghosh}
\orcid{0000-0003-1637-267X}{Kuntal}{Misra}
\orcid{0000-0001-9868-9042}{Dimple} {}
\end{orcids}

\begin{authorcontributions}
All authors in this work have made significant contributions.
\end{authorcontributions}

\begin{conflictsofinterest}
The authors declare no conflict of interest.
\end{conflictsofinterest}

\end{furtherinformation}

\bibliographystyle{bullsrsl-en}

\bibliography{extra}

\end{document}